\documentclass[10pt,preprint]{aastex}

\newcommand{\CIV}{\hbox{{\rm C}\kern 0.1em{\sc iv}}}
\newcommand{\MgII}{\hbox{{\rm Mg}\kern 0.1em{\sc ii}}}
\newcommand{\HI}{\hbox{{\rm H}\kern 0.1em{\sc i}}}
\newcommand{\HII}{\hbox{{\rm H}\kern 0.1em{\sc ii}}}
\newcommand{\Ly}{\hbox{{\rm Ly}\kern 0.1em$\alpha$}}
\newcommand{\Ha}{\hbox{{\rm H}\kern 0.1em$\alpha$}}
\newcommand{\fdlam}{erg~s$^{-1}$~cm$^{-2}$~\AA$^{-1}$}
\newcommand{\flux}{erg~s$^{-1}$~cm$^{-2}$}

\newcommand{\mpy}{M$_{\odot}$~yr$^{-1}$}
\newcommand{\lsun}{L$_{\odot}$}
\newcommand{\cmsq}{\hbox{cm$^{-2}$}}

\begin{document}

\title{H$\alpha$ Imaging with {\it Hubble Space Telescope}-NICMOS of an Elusive Damped Ly$\alpha$ Cloud at $z$ = 0.6~\footnote{
Based on observations with the NASA/ESA {\it Hubble Space Telescope} obtained at
the Space Telescope Science Institute, which is operated by Association of
Universities for Research in Astronomy, Inc. under NASA contract NAS5-26555}
}
\author{Nicolas Bouch\'e } 
\affil{Department of Astronomy, University of Massachusetts, LGRT 531, Amherst, MA 01003; bouche@nova.astro.umass.edu}
\author{James D. Lowenthal}
\affil{Department of Astronomy, University of Massachusetts, LGRT 526, Amherst, MA 01003; james@astro.umass.edu}
\author{Jane C. Charlton}
\affil{Department of Astronomy and Astrophysics, Pennsylvania State University, University Park, PA 16802; charlton@astro.psu.edu}
\author{Matthew A. Bershady}
\affil{Department of Astronomy, University of Wisconsin, Madison, 475 North Charter Street, Madison, WI 53076-1582; mab@astro.wisc.edu}
\author{Christopher W. Churchill}
\affil{Department of Astronomy and Astrophysics, Pennsylvania State University, University Park, PA 16802; cwc@astro.psu.edu}
\and
\author{Charles C. Steidel}
\affil{California Institute of Technology, Palomar Observatories, Pasadena, CA 91125; ccs@astro.caltech.edu}

\begin{abstract}
Despite previous intensive ground--based imaging and spectroscopic
campaigns and wide-band HST
imaging of the $z=0.927$ QSO 3C336 field, the galaxy that hosts 
the damped \Ly\ system along this line--of--sight has eluded detection.
We present a deep narrow-band \Ha\ image of the field of this
$z_{abs}=0.656$ damped \Ly\ absorber, obtained through the
F108N filter of NICMOS 1 onboard the {\it Hubble Space Telescope}.
The goal of this project was to detect {\it any} \Ha\ emission
10 times closer than previous studies to unveil the damped absorber.
We do not detect \Ha\ emission between 0.05\arcsec\ and 6\arcsec\ 
(0.24 and 30 $h^{-1}$~kpc) from the QSO,
 with a 3$\sigma$ flux limit of $3.70\times 10^{-17} h^{-2}$ \flux\
 for an unresolved source, corresponding to a star formation rate (SFR)
 of $0.3 h^{-2}$~\mpy.
This leads to a 3$\sigma$ upper limit of 0.15 \mpy~kpc$^{-2}$ on the SFR density, 
or a maximum SFR of 1.87~\mpy\ assuming a disk of $4$~kpc in diameter.
This result adds to the number of low redshift damped \Ly\ absorbers
that are {\it not} associated with the central regions of Milky-Way-like 
disks.  Damped \Ly\ absorption can arise from high density concentrations 
in a variety of galactic environments including some that, despite their 
high local \HI\ densities, are not conducive to widespread star formation.

\end{abstract}


\section{Introduction}

QSO absorption lines provide a powerful  approach to studying the
history of galaxies.  The nature of damped \Ly\ absorbers (DLAs)
towards background QSOs has been an ongoing debate for more than a
decade.  \citet{wolfe86,wolfe95} proposed that DLAs are large
progenitors of today's massive spiral disks.  Evidence in support of
this interpretation includes the measurement of absorption line
velocity profiles that are consistent with those expected from lines
of sight intercepting rotating, thick gaseous disks \citep{prochaska}.
However, recent theoretical simulations of galaxy formation showed
that a large range in structures (e.g. halo gas clouds) and
morphologies, rather than a single uniform type of galaxy, can give
rise to DLAs (e.g. \citep{katz,haehnelt,mcdonald}).  In this case,
they could be low surface brightness (LSB), gas rich, dwarf galaxies
as proposed by \citet{tyson} and more recently by \citet{jimenez}.

DLAs make up the largest reservoir of neutral hydrogen (\HI) at high
redshift \citep{wolfe86,lanzetta,rao00}.  There are hundreds of DLAs,
defined as absorbers with atomic hydrogen column densities $N(\HI)$
greater than $ 2 \times 10^{20}$~{\cmsq}, known up to redshifts
$z\sim4$.  From the redshift distribution of the measured column
densities of the damped systems,
the evolution of neutral gas density ($\Omega_{DLA}(z)$) can be
measured \citep{lanzetta,lanzetta95,wolfe95,storrie96,rao00}.  The
analysis of the variation of $\Omega_{DLA}(z)$ with redshift can
provide another measurement of the evolution of the star formation
rate in the Universe \citep{pei}.  \citet{rao00} find little evolution
in $\Omega_{DLA}(z)$ from $z=4$ to $z=1$.  Over this interval we might
have expected a decrease in $\Omega_{DLA}$ as the gas is converted
into stars in order to maintain the observed constant star formation
rate \citep{steidel99}.  This led \citet{rao00} to conclude that DLAs
and the galaxies that dominate the star formation density are
different populations.

Furthermore, it is difficult to reconcile the low metallicities of
high-redshift DLAs (typically 1/10 of solar at $z\sim2.5$,
\citet{pettini}) with the higher metallicities of stars in galaxies
today: no chemical evolution is seen in DLAs from $z=3.5$ to $z=0.3$
\citep{pettini99}, which indicates that DLAs do not necessarily trace
the population responsible for the bulk of star formation.  Finally, high
resolution Keck spectra of three QSOs by \citet{pettini00} indicate
that DLAs ($z<1.0$) have heterogeneous chemical properties.  

Another approach to understanding DLAs is to compare them with current
local \HI\ surveys \citep{briggs,zwaan,rosenberg}, which sample
neutral gas clouds and galaxies perhaps analogous to distant DLAs (of
course, high redshift and low redshift DLAs may be produced by very
different galaxy populations).  Current results \citep{zwaan,rao00}
indicate that local \HI\ samples contain a much smaller fraction of
high column densities ($N_{\HI} >10^{21}$~cm~$^{-2}$) than both low and
high redshift DLAs.  These results imply strong evolution at the
highest column densities.  However, estimates of the local column
density distribution rely on many caveats, such as the true nature of
the debated local \HI\ mass function.  Beam-smearing in these \HI\
surveys may also lead to underestimates of the local number of high
column density systems, particularly if such regions are physically
small.

Prior to 1993, most observations of DLAs were restricted to high
redshift systems ($z>1.8$) since the \Ly\ line is in the rest--frame
UV.  Previous attempts to detect emission from DLAs have concentrated
on \Ly\
%(e.g., Lowenthal et al. 1995; see Roche, Lowenthal,\& Woodgate 2000 for a summary),
(e.g., \citet{lowenthal95}; see \citet{roche} for a summary), which is
expected to be a signature of a star-forming region, although its
emission may be suppressed by dust extinction.  Of at least ten DLAs
at $z\ga 2$ searched for \Ly\ emission, only a very few show confirmed
detections \citep{djorgovski,fynbo00}.  Ground based surveys
(photometric and spectroscopic) for {\Ha} emission around $z>2$ DLAs
have also been mostly unsuccessful \citep{bunker,teplitz}, except in some
cases \citep{bechtold,mannucci}.

Only recently, with the advent of rest-UV spectroscopy from {\it
Hubble Space Telescope (HST)}, have data on intermediate redshift
$0.3<z<1.8$ DLA systems become available
\citep{lebrun,boisse}.  These DLAs display a wide range of morphologies
and surface brightnesses.  Using {\it NICMOS-NIC2}, \citet{kulkarni}
reported a possible {\Ha} detection of a $z_{abs}=1.89$ DLA towards LBQS 1210+1731
and \citet{pettini00} detected an edge-on low luminosity $L\sim 1/6
L^*$ galaxy 10~kpc away from the QSO 0058+019 ($z_{abs}=0.612$) 
using {\it WFPC2}.  In
addition, ground based observations such as WIYN images of three low
redshift DLAs ($z < 0.3$) show dwarf and/or low surface brightness
hosts, with confirmed redshifts \citep{rao98,lane98}.

In contrast, the larger class of {\MgII} absorbers with $3\times
10^{17} \leq N({\HI})\leq 2\times 10^{20}$~{\cmsq} (Lyman limit
systems) are almost always associated with fairly luminous
($L_K>0.05L_K^*$) galaxies, i.e. within $35 h^{-1}$~kpc
\citep{steidel95}.  It was once thought that these different classes
of absorbers sample different cross sectional regions of broadly the
same galaxies \citep{steidel93}, with the DLAs associated with the
inner, denser regions.  This is only partially true.  Dwarf and LSB
galaxies apparently can produce DLA absorption at low impact
parameters; however, they do not contribute a significant
cross--section for Lyman limit absorption.

These issues motivate the current attempts to image DLAs at low impact
parameter and, at the same time, might explain why most previous
attempts have failed to reveal gas-rich spirals.

In this paper, we present an {\it HST} study of an intermediate
redshift DLA ($z_{abs}=0.656$) along the line-of-sight towards QSO
3C336 ($z$=0.927) that has, so far, eluded detection entirely despite
extensive ground-based searches \citep{steidel}.  This QSO
line-of-sight is one of the richest known for $z<1$ absorption line
studies, with $6$ metal line systems in the interval $0.317 < z <
0.892$.  For that reason this quasar field was the target of both a
very deep $24000$~s {\it HST--WFPC2} image and a $2160$~s Keck/NIRC
image \citep{steidel}.  Five galaxies associated with the metal line
systems were identified in the WFPC2 image and their redshifts
confirmed spectroscopically using the Keck/LRIS \citep{steidel}.  The
only unidentified absorber is a DLA at $z=0.656$ with $N(\HI)=2\times
10^{20}$~{\cmsq} and $[Fe/H]=-1.2$.  There is no galaxy detected
with $L>0.05L_K^*$ near the QSO line-of-sight 
 and as close as 0.5\arcsec ($\sim 2 h^{-1}$~kpc).  Two unlikely candidate
galaxies exist at large impact parameters from the QSO. The first is a
relatively faint, $M_K =-21.43$ ($m_K=20.77$), late type spiral
located 14.3{\arcsec} ($\sim 65 h^{-1}$~kpc) NE of the QSO.  Taking
into account the estimated disk inclination and the position angle,
\citet{steidel} estimated it would require a disk extent of at least
$\sim 120 h^{-1}$~kpc to intercept the QSO line--of--sight.  The second
candidate is a galaxy without a confirmed redshift.  If one assumes
this galaxy is at the redshift of the DLA, it would have an impact
parameter of $41 h^{-1}$~kpc and $L_B=0.04L_B^*$ (similar to that of the
SMC).

This raises the question of whether DLA absorption can arise in dense
\HI\ regions far from the centers of galaxies, perhaps in regions that
have little or no current star formation, as seen locally in mergers
\citep{hibb99} and as pointed out by \citet{rao00}.  Alternatively, a
separate absorbing galaxy could be situated beneath the QSO on the
plane of the sky.

The goal of this project was to detect {\it any} \Ha\ emission as close as
$\sim 0.05$\arcsec ($0.24~h^{-1}$~kpc) of the QSO to test
further the alternate hypothesis.  
\Ha\ at the redshift of this DLA matches one narrow--band filter of the {\it HST-NICMOS} camera~1 
 and, therefore,  enables us to put strong constraints on the SFR of the absorber.
We can already say there is no $L^*$ spiral galaxy close to the line-of-sight,
for \citet{steidel} did not find anything brighter than $0.05 L_K^*$.
On the other hand,
one might expect a dwarf or LSB galaxy with significant star formation 
such as found by \citet{lebrun} and \citet{rao00} for other DLAs.

In the next section, we describe the observations and the reduction of
the data.  The results are given in section \ref{sec_results}, and
we compare them with previous studies in section \ref{discussion}.
Our conclusions are presented in section \ref{conclusions}.
Throughout this paper, we adopt
$\Omega_M=0.3$, $\Omega_{\Lambda}=0.7$, and $H_0=100 h$ km s$^{-1}$~Mpc$^{-1}$;
thus 1\arcsec\ at $z=0.656$ corresponds to $4.85 h^{-1}$~kpc.

\section{Observations and Data Reduction}

\subsection{Observations}

The observations were carried out with the Near Infrared Camera and
Multi-Object Spectrometer (NICMOS) using Camera 1 (NIC1) with Camera 2
(NIC2) in Attached Parallel mode on 1997 October 3.  NIC1
(0.043\arcsec~pix$^{-1}$) was chosen to enable us to over-sample the
point spread function (PSF) (FWHM $\sim 2$~pix or 0.09\arcsec) and 
hence to resolve emission as close as
possible to the QSO.  The exposure times were 2.8hr (5 orbits) and
0.7hr for the filters F108N  and F110M respectively (details are listed in
Table~\ref{expos}).  The pixel size is 0.043\arcsec\ 
 which at the redshift  of the DLA
corresponds to a physical size of $\sim 0.21h^{-1}$ kpc~pixel$^{-1}$.

For NIC1, we used the narrow--band  (NB) F108N filter ($\lambda_c
=1.081\mu$m; $\mbox{FWHM}=0.0094 \mu$m) and the broad--band (BB) F110M filter 
($\lambda_c=1.101 \mu$m; FWHM$=0.19\mu$m).
 F108N was chosen to match
the wavelength of \Ha\ at the redshift of the DLA ($z=0.656$) and
the BB filter was chosen to subtract continuum emission.
For NIC2, the filter F160W ($\lambda_c=1.596 \mu$m; FWHM$=0.400\mu$m) was 
chosen  but unfortunately, the field turned out to be in an
especially empty region of the WFPC2 field \citep{steidel}.
The NIC2 images will not be discussed further here.

A reference star (GSC2044.810 from the STScI Guide Star Catalog) to
measure the PSF was chosen within 100\arcsec\
of QSO 3C336. GSC2044.810 has colors $J-H\sim0.6$ and $H-K\sim 0.6$, 
which are similar to that of a typical QSO at $z\sim1$, i.e.
 $B-V$=1.05 and $V-J$=1.68 \citep{hyland82}.
The star  ($V$=12.59) is $\sim 5.5$mag brighter  than the QSO,
 which allowed us to shorten exposure times.

Five dithered images of both the QSO and the reference star were taken
in both NB and BB filters in order to improve the sampling of the PSF
and to avoid any systematic noise from the detector. Each exposure was
taken in the MULTIACCUM mode, which allows 25 non-destructive readouts
over the entire dynamic range for each pixel. This allows improved
cosmic ray rejection. Exposures were chosen to be short enough to 
avoid detector saturation.

\subsection{Data reduction}

The images of the QSO and the  reference star
were reduced, shifted and coadded with the
Nicred\footnote{see http://cfa-www.harvard.edu/castles/Nicred for more
information.} package \citep{mcleod}.
The zeroth read was
subtracted from each raw image to remove any pedestal level, and dark
subtraction was performed using the pipeline dark files provided by
STScI.  The photon arrival rate in each pixel was computed by fitting
a straight line to the MULTIACCUM readouts.  Cosmic rays were rejected
by searching for a jump between successive readouts so that full
information for each pixel is recovered.  The images were then
flat-fielded with the pipeline flat files provided by STScI.
A sky frame was constructed from the dithered observations and
subtracted from the images.  Each image was magnified by a factor of 2
after masking the residual cosmic rays and bad pixels. Finally, the
dithered images were registered and coadded
-- with a weight proportional to  the inverse noise in a reference region 
close to the QSO image-- to produce the final image
(hereafter "high SNR" images).
The final result improved the SNR by a factor of $\sim 8$ compared to images
reduced by the STScI pipeline {\it calnica} process.  The FWHM of the
PSF is 4.17 pixels (in the magnified images), which corresponds to
0.090\arcsec.

In order to perform absolute photometry on the QSO flux, i.e. to use
the multiplying factors (PHOTFLAM) provided by STScI and to convert
counts to flux units, we reduced the data using the dark and flat
files listed as the Reference Files (used in the calibration of the
P330E and G191-B2B 7691 data) prescribed by the NICMOS handbook
v.3.0 (hereafter "calibrated" images). 
To check for any
correlation noise produced by the magnification process, we reduced
the data without magnifying the pixels (hereafter "small calibrated" images)
 with the same dark and flat files and
we confirmed our noise figures (measured around the QSO) in the calibrated images.
In the NB QSO field, the SNR was $\sim 125$ and $\sim 15$, 
respectively for the  high SNR, 
and  calnica images. In the  small calibrated and  calibrated
images, the SNR was $\sim 95$.
We used the high SNR images for PSF subtraction, while the noise properties
were measured in the   calibrated images. The calibrated image is shown 
in Figure~1.

\subsection{Photometry and noise properties.}

The total fluxes were obtained
by measuring the curve of growth for each of the calibrated QSO
images. We converted the fluxes measured within 0.5\arcsec\ radius
aperture to nominal infinite aperture fluxes by multiplying them by
1.15 as prescribed by STScI~\footnote{see
http://www.stsci.edu/cgi-bin/nicmos/
under {\it documents} and {\it handbook}. See also Figures 4.6 to 4.10 from the NICMOS Instrument Handbook v3.0 Chapter 4.}.

In the NB calibrated image, the noise beyond the PSF is
constant within a 2\arcsec--radius circle (from 10 to 50 pixels)
around the QSO. Beyond a radius of 2\arcsec, the noise increases
due to the poorer first quadrant of the NICMOS detector.

Our 3-$\sigma$ detection limit for a point source is given by the
3-$\sigma$ rms per resel (a resel or resolution element is a
0.09\arcsec\ diameter aperture, corresponding to the FWHM of the PSF)
measured in the quadrant of the image that includes the QSO.  For the
NB image, our 3-$\sigma$ detection limit is: $F_{\lambda} = 3.78
\times 10^{-19}$ \fdlam ($m_{AB}= 23.48$, where $m_{AB}=-2.5\times
\log (F_{\nu})-48.6$), or a flux of $F_{\Ha} = 3.70 \times 10^{-17}$ \flux. For
the BB image, our 3-$\sigma$ detection limit is: $F_{\lambda} = 2.9
\times 10^{-20}$ \fdlam ($m_{AB}=26.22$), or a flux 
of $F_{BB} = 5.82 \times 10^{-17}$ erg~s$^{-1}$~{\cmsq}.

Our 3-$\sigma$ detection limits for an extended source are given by the
3-$\sigma$ rms per pixel scaled by the square root of the number of pixels in
a 1\arcsec\ by 1\arcsec\ square.  The resulting limits are
$4.70\times 10^{-18}$ \fdlam ($m_{AB}=20.75$) in the NB image and
$3.60\times 10^{-19}$ \fdlam ($m_{AB}=23.49$) in the BB image.
These correspond to surface brightnesses of
$4.60 \times 10^{-16}$ \flux~arcsec$^{-2}$ for the NB image and
$7.17 \times 10^{-16}$ \flux~arcsec$^{-2}$ for the BB image.

The total flux densities of the QSO in the NB and the BB
are $(4.55 \pm 0.18) \times 10^{-17}$ \fdlam ($m_{AB}=18.3$)
and  $ (5.09 \pm 0.17) \times 10^{-17}$ \fdlam ($m_{AB}=18.1$),
respectively. These results are summarized in Table~\ref{results}.

\subsection{Profile subtraction}
In order to reveal any faint object with small impact parameter,
we subtracted the QSO PSF in the following ways:

To search for faint emission both close to the QSO
line of sight and throughout the 11\arcsec\ field,
we first subtracted the BB QSO image (scaled to the peak value) from
the NB QSO image.
The central part, shown on Figure~2 (a),
 has faint residuals (negative and positive peaks well below $3\sigma$)
 near the QSO position.  The total residual flux measured in
a  resel  centered on the QSO is
 about $1.5 \sigma$ above the mean or 0.7\% of
the (unsubtracted) NB QSO flux in the same aperture.

To assess how much of those residuals might be due to differences
in PSF between the F108N and F110M filters and how much to real \Ha\
flux from a dwarf galaxy exactly superposed on the QSO position, we
used the same procedure (i.e. shifting \& centering) to subtract the
BB PSF from the NB PSF of the reference star (GSC2044.810).  We find
residuals with a similar pattern (see Figure~2 (b)).
There is clearly a peak near the center with a depression above and
below (in the $y$-direction).  The stellar PSF residual pattern on larger
scales is not seen in the QSO PSF subtracted image, since the SNR is
more than ten times lower. 
 The central peak pixel in the residuals is about 1.5\% the flux
of the stellar PSF peak, while the magnitude of the deepest depression
is about -5\% of the stellar PSF peak.  The total residual flux
measured in a resel centered on
the position of the star is 0.7\% of the  NB stellar  PSF flux in the
same aperture, consistent with the QSO residuals.
Therefore, we conclude the residuals are likely due
 to PSF differences between the two filters.

Because the stellar NB - BB method described above produced such
significant residuals, we then tried subtracting a stellar PSF (scaled
to the peak) directly from the QSO high SNR image.  This was performed
for both the NB and BB filters.  We used both a theoretical PSF
generated by the software Tiny Tim (\citet{krist}\footnote{
 The program and informations are available online at http://www.stsci.edu/software/tinytim }
; adapted by Richard
Hook for NICMOS) and the reference star (GSC2044.810) PSF.
Unfortunately, the reference star turns out to be double, i.e. after
the subtraction, a PSF-like hole was seen in both NB and BB images
offset by 0.17\arcsec\ from the central PSF (see
Figure~3~(a)).  We corrected for this by subtracting
the primary component of the star PSF from the secondary component,
and then subtracted the result from the QSO PSF.  The final result
shows very little residual (see Figure~3~(b)).
However, when using the theoretical PSF
instead, the result tends to leave a ring-like structure. For this
reason, we adopted the stellar PSF as the best approximation of the
true PSF.  The final PSF subtracted NB image is shown in
Figure~3~(b) after a gaussian smoothing of 1.5 pixels.

Finally, in the NB PSF subtracted image, we looked for faint emission
throughout the 11\arcsec\ field both by eye and using the algorithm
SExtractor \citep{bertin} with a 3-$\sigma$ threshold (with a minimum
of 16 pixels above threshold, i.e. the keyword MINAREA in
SExtractor). Three candidates were found.  Every candidate was
followed up by examining individual images and  it turned out that all
candidates were artifacts left over from the reduction process,
e.g. cosmic ray residuals smeared out over several pixels by the
magnification process.

\section{Results and Discussion}
\label{sec_results}

To summarize, we looked for emission objects near the QSO and throughout the field
in the following way:
(i) We subtracted the BB image from the NB image. 
Apart from residuals due to PSF differences between the two filters, 
no emission was detected;
(ii) We subtracted the star PSF from the QSO PSF in both NB and BB images.
No emission was detected;
(iii) We searched by eye for any faint emission throughout
the 11\arcsec\ field in both the NB  PSF subtracted and the 'NB minus BB' images;
(iv) In the NB PSF subtracted image, we
also used the algorithm SExtractor \citep{bertin} to look for faint
emission. All three candidates were cosmic rays residuals smeared out over several pixels.
 Therefore, we conclude
that no emission objects were detected in either the NB  PSF subtracted,
 or in the 'NB minus NB' image.

We can use the lack of detection in the BB image to constrain the
presence of luminous galaxies.  The BB 3-$\sigma$ detection limit for
a point source is $m_{AB}=26.22$, which corresponds to a continuum
luminosity of $L_R=5.02 \times 10^{40} h^{-2}$ erg~s$^{-1}$
in the rest frame $R$--band or  $L_R=7.23 \times 10^{7} h^{-2}$ \lsun.

In addition, our lack of detection in the NB constrains
directly the SFR.
For the NB filter, our 3-$\sigma$ detection
limit for a point source (i.e. unresolved) corresponds to an \Ha\ luminosity of
$3.20 \times 10^{40} h^{-2}$ erg~s$^{-1}$ at the redshift of the DLA.
The 3-$\sigma$ detection limit for an extended source corresponds to
an \Ha\ luminosity of
$3.98 \times 10^{41} h^{-2}$ erg~s$^{-1}$arcsec$^{-2}$ 
at the redshift of the DLA (see Table~\ref{results}).  
Using the \citet{kennicutt} conversion factor for a constant SFR and a modified 
Salpeter--like IMF with variable slope, i.e. 
$SFR=L_{\mbox{\Ha}}/1.12\times10^{41}$ erg~s$^{-1}$, we derive
a SFR of $<0.28 h^{-2}$\mpy\ for an unresolved source or 
$<0.15$~\mpy~kpc$^{-2}$ for an extended source. 
Assuming the absorber is a disk of radius 2~kpc
(i.e. $\sim 9$ times the resolution element; see discussion below),
 this gives a SFR of $<1.87$ \mpy.

More recently, the calibration of \citet{kennicutt98}
 yields a similar conversion factor
$SFR=L_{\mbox{\Ha}}/1.26\times 10^{41}$ erg~s$^{-1}$.
Our upper limit is a conservative one given that the actual SNR in the
magnified images used in the subtraction process is higher than the
calibrated images and that the conversion factor is the smallest of
current estimates.  The SFR estimate is also dependent on the
IMF. Using a Salpeter IMF with a higher mass cutoff (125 M$_{\odot}$),
\citet{alonso} found $SFR=L_{\mbox{\Ha}}/3.1\times 10^{41}$
erg~s$^{-1}$, which is about 3 times larger than the \citet{kennicutt}
result.  This would {\it decrease} our upper limit by a factor of
three, i.e. a $SFR < 0.6$ \mpy\ for a 2~kpc-radius disk.

This estimate depends strongly on the assumed size of the
object.  However, from \citet{steidel}, there can not be any $L>0.05
L^*$ galaxy (typically 10~kpc in size) as close as 
0.5\arcsec($2h^{-1}$~kpc).
In other words, anything larger than 4~kpc (in diameter) would
have been seen in both \citet{steidel} and in our images.  A
2~kpc-radius object is consistent with the size of the
DLA candidate ($z=1.89$) of \citet{kulkarni}.
% DLA absorber ($z=1.89$) observed by \citet{kulkarni}.
%The compact objects
%reported by Le Brun et al. (1997; see section~4) have sizes of $1$--$3$~kpc.
Similarly,  Le Brun et al. (1997; see section~4) detect compact objects
with sizes of $1$--$3$~kpc along DLA lines-of-sight.
On the other hand, if the absorbing object is smaller, it could have a higher
SFR.  This would require that it be smaller than 1 resel or $\sim
0.5$~kpc, and that it be exactly aligned with the line of sight
of the QSO.

\section{Comparison with previous studies}
\label{discussion}

We first discuss our result in the context of previous surveys of
DLAs, and then compare our SFR limits to four types of galaxy seen
in the local universe.

From {\it HST} imaging of the fields of seven quasars with DLAs at
$0.395<z<1.78$, \citet{lebrun} were able to resolve galaxy--like
objects at small impact parameters for six of their QSO lines-of-sight
with $0.395<z<1.1$.  However, the hosts of these DLAs displayed a
broad range of morphologies and surface brightnesses: three of the six
detections are spirals, two are compact, LSB galaxies, and two others
have compact morphologies.  Recently, \citet{pettini00} reported the
detection of an edge-on low luminosity $M_B=-19.1$ or $L\sim 1/6 L^*$
galaxy 10~kpc away from the QSO 0058+019 using {\it HST--WFPC2}.
Using NICMOS-NIC2 \citet{kulkarni} reported a possible detection of an \Ha\
emission feature, $2$--$3$ $h^{-1}_{70}$~kpc in size, $0.25$\arcsec\
from a $z=1.89$ DLA.  They suggest that a faint, compact, somewhat
clumpy object, rather than a thick, spiral disk, is responsible for
this DLA.  The implied 3-$\sigma$ upper limit on the SFR is
4$h^{-2}_{70}$\mpy, which applies unless dust obscuration is
important.  Note that we used the same conversion factor
($L_{\mbox{\Ha}}/SFR$) as \citet{kulkarni}.

The ground-based spectroscopic survey of \citet{bunker}, which
searched for redshifted \Ha\ emission in 11\arcsec x2.5\arcsec\
regions around 6 quasars with $z>2$ DLAs, reached a 3-$\sigma$
detection limit of 6--18~{\mpy} and failed to detect any redshifted
\Ha\ emission.  
Some ground-based narrow-band photometric surveys (e.g., Teplitz, Malkan \& McLean 1998)
for \Ha\ emission from DLAs have also failed to detect any emission
line objects in the DLA fields, although \citet{teplitz} found \Ha\
emitters in the fields of some weaker non-DLA metal line systems.
However, other narrow-band searches for \Ha\ emission have revealed
multiple objects in the DLA fields separated by more than several to
tens of arcseconds from the QSO (\citet{bechtold}; \citet{mannucci}).
These surveys, which had 3-$\sigma$ detection limits of $\sim
5$--$10$~\mpy, found these \Ha\ emitting objects to have a wide range
of SFRs, $6$--$90$~\mpy.  \citet{kulkarni} suggested the relatively
large separations of these emission line objects from the quasar
indicate that they are not the DLA absorbers themselves, but
star-forming regions in a group or cluster also containing the DLA.
None of these ground-based surveys has been able to probe the regions
very close ($<2\arcsec$ or $11.7$~kpc at $z=2$) to the quasar
line-of-sight to rule out large spiral disks at high redshift with
confidence.

The nature of the DLA towards 3C336 can be addressed by comparing
directly our SFR upper limit to various types of galaxy in the local universe.
(i) {\it LSB  galaxies} have
{\HI} surface densities of $\sim 5$ M$_{\odot}$~pc$^{-2}$ 
or $N_{\HI}\sim 6.5 \times 10^{20}$~{\cmsq} \citep{vanderhulst}, well above
the DLA cutoff of $2 \times 10^{20}$~{\cmsq} but below the value of
$10^{21}$~{\cmsq} usually quoted as the threshold for star formation.
The mean star formation rates, derived from the \Ha\ luminosities,
in LSBs are typically $\sim0.1$ \mpy\ \citep{vandenhoek}.
(ii) {\it Blue compact dwarf galaxies (BCDG)}:
From CFHT observations, \citet{petrosian} reported a SFR of  
$0.3$--$0.5$~\mpy~kpc$^{-2}$ for the two main
\Ha\ emitting regions of the local ($D=10$~Mpc) BCDG, IZW 18;
(iii) {\it Typical spiral galaxies}: The typical SFR for a spiral
galaxy is $\sim 10$ \mpy\ \citep{vandenhoek}, spread over at least several kpc;
(iv) {\it Starbursts} can have a much larger SFR,
e.g. Arp 220  is forming stars at a rate of $\sim 240$\mpy, 
derived from {\Ly} \citep{anantharamaiah}.
The SFR in the DLA at $z=0.656$ toward 3C336 is far less extreme 
than in a starburst or a typical spiral, and somewhat less extreme 
than a BCDG.  However, it is consistent with that of an LSB galaxy
and it could be even lower, i.e. zero.

Another example of a region of neutral hydrogen that exceeds the DLA
threshold, but does not have significant star formation, has been
found through 21-cm observations of the giant \HI\ cloud 1225+0146
\citep{gh89}; these data show that neutral gas structures with little or no
star formation do exist.  This suggests that factors beyond a simple
column density threshold govern the formation of stars.  This giant
\HI\ cloud is 200~kpc along its major axis and has two peaks of {\HI}
emission, with $N(\HI) = 2 \times 10^{20}$ and $1 \times
10^{20}$~{\cmsq}, separated by 100~kpc.  In a deep optical search,
only a faint $M_B = -15.5$ dwarf irregular, 5~kpc in extent, was
discovered, corresponding to the largest peak in emission
\citep{mcmahon}.  Furthermore, no galaxy has been detected near the
$10^{20}$~{\cmsq} peak, which is almost enough neutral hydrogen to
produce a DLA.

The absorption line properties of the DLA towards 3C336 corroborate
the observed low SFR.  This DLA is unusual in that it is a
``{\CIV}--deficient'' {\MgII} absorber --- rest frame equivalent width 
$W({\CIV})\sim 0.5$\AA\
\citep{steidel}---, i.e. $W({\CIV})$ is less than half the typical
value for DLAs.  \citet{cwc99} found a correlation between the
strength of {\CIV} absorption and the kinematic spread of the {\MgII}
profile in high resolution absorption profiles, with the exception of
several {\CIV}--deficient absorbers.  They hypothesize that there is a
relationship between the strength of {\CIV} absorption (which is
generally consistent with arising in a corona similar to that around
the Milky Way disk) and the level of star formation activity in the
disk.  The kinematic spread of the {\MgII} profile is also thought to
be related to star forming processes that either eject or are
triggered by high velocity {\MgII} clouds.  In this scenario, the
{\CIV} deficient absorbers would be characterized by a lower than
average star formation rate, and in fact a few of them do have red
colors \citep{cwc00}, rather than the blue colors of actively
star-forming systems.  This is consistent with the fact that the 3C336
DLA has both a small $W({\CIV})$ and a strong limit on the star
formation rate in its vicinity.

There is at least some theoretical reason to expect low SFRs from
DLAs.  \citet{mo} hypothesize that --- at least at $z\sim 3$ --- 
Lyman break galaxies (LBGs), which are selected partly by their
strong star formation, and DLAs could be disjoint populations: If LBGs are the
central galaxies of massive dark halos at $z\sim 3$, then they should
be small objects with substantial star formation but low angular
momentum, while DLAs should be biased towards objects with large
angular momentum.  In the hierarchical structure formation models of
\citet{maller}, meanwhile, DLAs arise from the combined effects of
massive central galaxies and a number of smaller satellite galaxies in
virialized halos, rather than only the central galaxies, so predicted
SFRs associated with DLAs are low.  Another interesting result
from Maller et al.'s models is that the impact parameter distribution has a
longer tail at $z\sim 1$ (up to 150~kpc) than at redshift $z\sim 3$,
which could then reconcile the observed galaxy 120~kpc away from the
line-of-sight of 3C336 and the $z=0.656$ absorber.

\section{Summary and Conclusions}
\label{conclusions}

We can summarize previous observational results on DLAs as follows:
(i) DLAs show
little evolution of $\Omega(z)$ from $z=4$ to $z=1$ \citep{rao00};
(ii) DLAs show little metallicity evolution \citep{pettini99} from
$z=3$ to $z=0.3$; (iii) $z<1$ DLAs have heterogenous chemical
properties \citep{pettini00}; (iv) DLAs have various morphologies
\citep{lebrun,kulkarni,rao00} (although it is possible that at least
{\it some} DLAs are large disks like the galaxies found by
\citet{lebrun} and \citet{pettini00}).  These results are inconsistent with the
standard, DLA/{\HI} disk paradigm\citep{wolfe95} for they indicate
that DLAs (1) do not participate in the overall chemical enrichment of
the universe and hence do not trace star formation; and (2) are not
characteristic of a particular type of galaxy.  Rather they are merely
characteristic of a particular type of region: namely one with a large
neutral column density (e.g. \citet{kher96}).  It is quite reasonable,
following the results of  \citet{rao00}, to assume that this applies to
higher redshift DLAs as well.

Nearly all space-based observations of low redshift DLAs have revealed
star formation in some sort of galaxy.  The $z=0.656$ system toward
3C336 is an exception.  Our lack of detection of any source with SFR
greater than 0.28$h^{-2}$~\mpy, or $0.15$~\mpy~kpc$^{-2}$,
demonstrates that DLAs with very little star formation can exist.

As evidence mounts against the standard, DLA/{\HI} disk scenario,
we must address the question: ``What is a DLA?''.
Possibilities not yet ruled out include knots of {\HI} hundreds of kpc
away from the main galaxy as seen locally \citep{hibb99}; small
compact dwarfs or LSBs; or photo-dissociation regions that may produce
much of the {\HI} currently observed in galaxy disks \citep{smith}.
In some cases, both a cold (hundreds of K) and a warm (thousands of K)
neutral medium are found along the line of sight through a DLA, based
upon analysis of 21--cm emission profiles \citep{lane}, while in
others the warm phase dominates.
   
Our limits on \Ha\ emission from the DLA towards 3C336 set the tightest
constraints yet on star formation in a compact absorber at
intermediate redshift, and probe closer to the QSO line-of-sight than
any previous study.  Our non-detection adds to the mounting evidence
that low redshift DLAs are made of galaxies of diverse morphologies,
luminosities and surface brightnesses rather than a uniform
population of luminous disk galaxies.

\acknowledgments
Support for this work was provided by NASA through grant number
GO-07449.01-96A from the Space Telescope Science Institute, which is
operated by the Association of Universities for Research in Astronomy,
Inc. under NASA contract NAS5-26555.  MAB acknowledges support from  
 NASA LTSA contract NAG5-6032.

\clearpage

\begin{figure}
\plotone{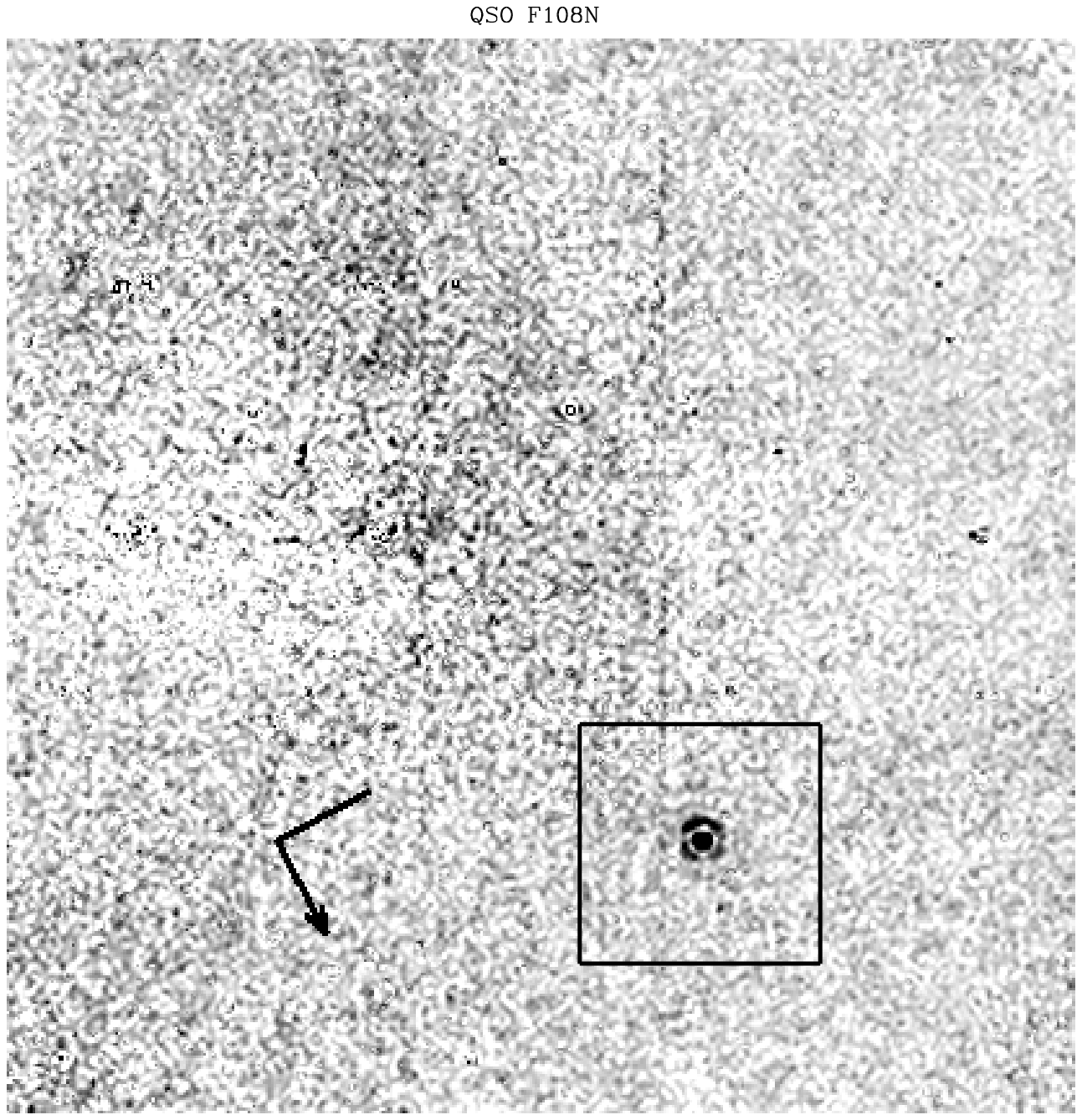}
\caption{
The calibrated NIC1 Narrow Band F108N image of the 11\arcsec x11\arcsec\ field around QSO 3C336.  The box is 2\arcsec x2\arcsec\ and centered on the QSO.
The arrow indicates North. The image is shown in reverse: dark is positive flux.
}
\end{figure}

%\figcaption[fig1.ps] {
%The calibrated NIC1 Narrow Band F108N image of the 11\arcsec x11\arcsec\ field around QSO 3C336.  The box is 2\arcsec x2\arcsec\ and centered on the QSO.
%The arrow indicates North. The image is shown in reverse: dark is positive flux.
%}

\begin{figure}
\plotone{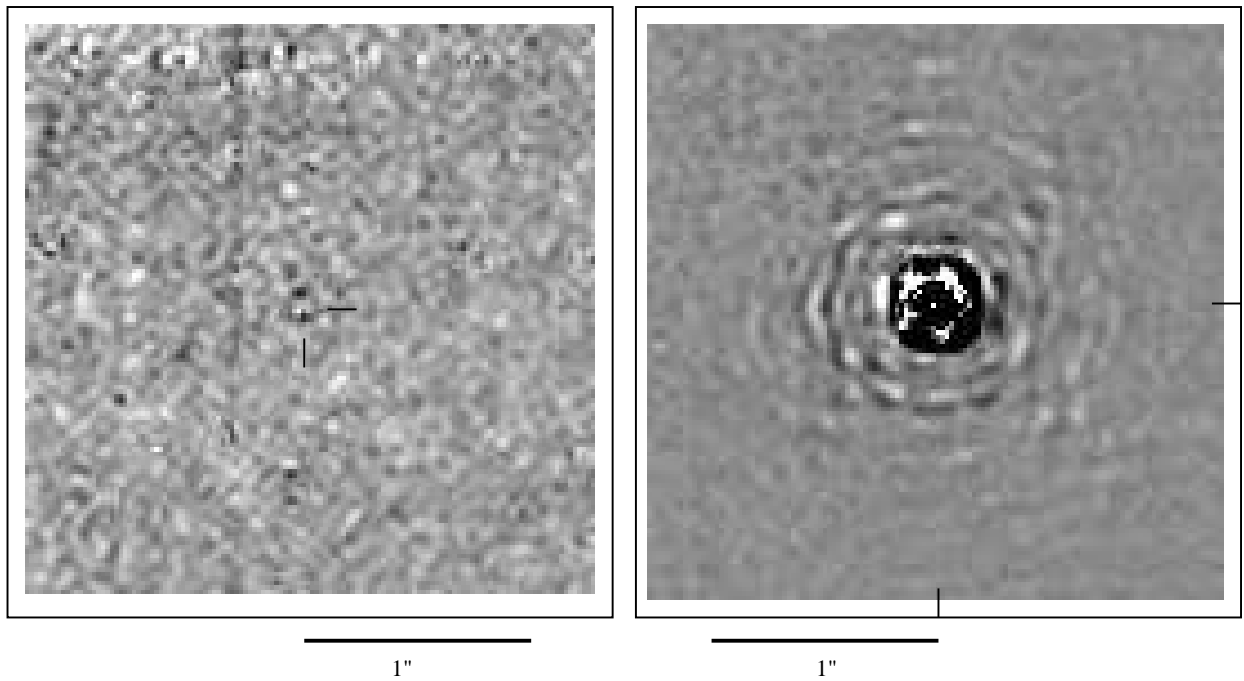}
\caption{
(a) Left: Narrow band NIC1 image of the QSO field after subtraction of
 broad band NIC1 image. 
Images were registered and scaled to the peak value prior to subtraction.
Residual flux
in 1 resel at the QSO position is only $1.5 \sigma$ above the mean.
The original position is shown by the tick marks.
(b) Right: same for the reference star.
Note the significant residual flux pattern, which is consistent with
the QSO residuals, although at much higher signal-to-noise ratio.
In both images, positive flux is shown as white. 
}
\end{figure}

%\figcaption[fig2.eps]{
%(a) Left: Narrow band NIC1 image of the QSO field after subtraction of
% broad band NIC1 image. 
%Images were registered and scaled to the peak value prior to subtraction.
%Residual flux
%in 1 resel at the QSO position is only $1.5 \sigma$ above the mean.
%The original position is shown by the tick marks.
%(b) Right: same for the reference star.
%Note the significant residual flux pattern, which is consistent with
%the QSO residuals, although at much higher signal-to-noise ratio.
%In both images, positive flux is shown as white. 
%}

\begin{figure}
\plotone{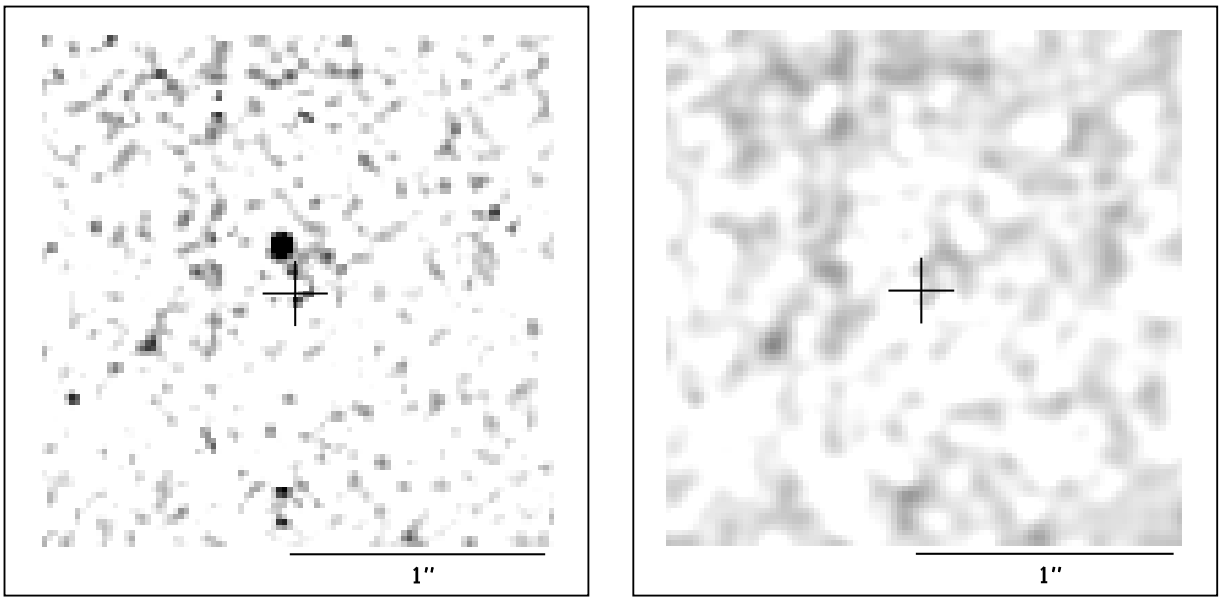}
\caption{
(a) Left: Blow up of 2\arcsec\ x2\arcsec\ around the QSO after subtraction of 
the scaled star PSF.  The cross marks the position of the QSO.
 The star turns out to be a close double,
resulting in a "hole" 0.17\arcsec\ from the QSO. 
(b) Right: Same as (a) after second PSF subtraction to remove
 the stellar companion.
 Image has been smoothed to 1.5 pixel resolution to
 enhance any extended emission. No residual flux is detected.
 (a) \& (b) are the "high SNR" images and  positive emission is
in white.
}
\end{figure}

%\figcaption[fig3.eps]{
%(a) Left: Blow up of 2\arcsec\ x2\arcsec\ around the QSO after subtraction of 
%the scaled star PSF.  The cross marks the position of the QSO.
% The star turns out to be a close double,
%resulting in a "hole" 0.17\arcsec\ from the QSO. 
%(b) Right: Same as (a) after second PSF subtraction to remove
% the stellar companion.
% Image has been smoothed to 1.5 pixel resolution to
% enhance any extended emission. No residual flux is detected.
% (a) \& (b) are the "high SNR" images and  positive emission is
%in white.
%}

\clearpage

\begin{deluxetable}{lrr}
\tablecaption{Exposure times for each filter.\label{expos}}
\tablewidth{0pt}
\tablehead{
\colhead{}  & \colhead{F108N} & \colhead{F110M} \\
\colhead{Object}  & \colhead{(s)} & \colhead{(s)} }

\startdata
3C336   &       $5\times 2050$s & $5\times 511$s \\
GSC2044.810 &   $5\times 303$s  & $6\times 23$s \\
GSC2044.810 &                   & $ 5 \times 14$s \\
\enddata
\end{deluxetable}

\clearpage

\begin{deluxetable}{ll}

\tablecaption{Summary of results in QSO 3C336 field.  \label{results}}
\tablecolumns{2}
\tablewidth{0pt}
\tablehead{
\colhead{Parameter} & \colhead{Value}  \\ \hline 
\multicolumn{2}{c} { QSO ($z$=0.927) } }
\startdata

R.A., decl. (J2000)			& $16^h24^m39^s.13, +23\arcdeg 45\arcmin 11.7\arcsec $ \\

$F_\lambda$ (1.08 $\mu$m)   (\fdlam)    &  $(4.55 \pm 0.18) \times 10^{-17}$ \\
$m_{AB,1.08 \mu \mbox{m}}$  (mag)       & $18.28 \pm 0.1$  \\
$F_\lambda$ (1.10  $\mu$m)  (\fdlam)    &  $(5.09 \pm 0.17) \times 10^{-17}$ \\
$m_{AB,1.10  \mu \mbox{m}}$ (mag)       & $18.11 \pm 0.1$  \\

\cutinhead{$3-\sigma$ Upper Limits on Continuum Emission at $z=0.656$}
$F_\lambda$ (1.10  $\mu$m)      (\fdlam)& $<2.92 \times 10^{-20}$  \\
$m_{AB,1.10  \mu \mbox{m}}$     (mag)   &  $<26.22$ \\
$F_{\mbox{BB}}$  (Unresolved source) (\flux)  &  $<5.82 \times 10^{-17}$ \\
$L_{\mbox{R}}$  (Unresolved source)  ($h^{-2}$ erg~s$^{-1}$)  &  $<5.02 \times 10^{40} $\\

\cutinhead{$3-\sigma$ Upper Limits on \Ha\ Emission from the Damped \Ly\ Cloud ($z=$0.656)
}
$F_\lambda$ (1.08 $\mu$m)     (\fdlam)  &  $<3.78 \times 10^{-19}$\\
$m_{AB,1.08 \mu \mbox{m}}$    (mag)     &  $<23.48$ \\
$F_{\mbox{\Ha}}$  (point source) (\flux) & $<3.70 \times 10^{-17}$  \\
$L_{\mbox{\Ha}}$(point source)   ($h^{-2}$ erg~s$^{-1}$)    & $<3.20\times 10^{40} $ \\
$\Rightarrow$ SFR (point source) ($h^{-2}$ \mpy)           &  $<0.28$ $h^{-2}$  \\
\\
$\mu_{\mbox{\Ha}}$ (3$\sigma$) (\flux arcsec$^{-2}$) \tablenotemark{a}   & $<4.60\times 10^{-16}$\\
$\Sigma_{\mbox{\Ha}}$ ($h^{-2}$ erg~s$^{-1}$  arcsec$^{-2}$) \tablenotemark{b}   & $< 3.98\times 10^{41}$\\
$\Sigma_{\mbox{\Ha}}$ (erg~s$^{-1}$kpc$^{-2}$) \tablenotemark{b}    & $<1.67\times 10^{40}$\\
$\Rightarrow$ Surface SFR   ( \mpy kpc$^{-2}$)                     & $<0.15$  \\
$\Rightarrow$ SFR (r=2kpc  disk) (\mpy)                     & $<1.87$  \\
\enddata
\tablenotetext{a}{\Ha\ surface brightness.}
\tablenotetext{b}{\Ha\ surface luminosity.}
\end{deluxetable}

\end{document}